\documentclass[iop,trackchanges,linenumbers]{emulateapj}
\usepackage{graphicx}
\usepackage{apjfonts}
\usepackage{amsmath} 
\usepackage{natbib}
\usepackage{color}
\usepackage[normalem]{ulem}
\DeclareOldFontCommand{\rm}{\normalfont\rmfamily}{\mathrm}
\newcommand{\prima}{^{\prime}}

\newcommand{\fdeg}{.\!\! ^{\circ}}

\def\Htildeel{\tilde{\rm H}}

\def\H{{\rm H}}

\def\degree{^{\circ}}

\def\PSF{{\cal S}}

\def\atilde{\tilde{a}}
\def\btilde{\tilde{b}}
\definecolor{al}{rgb}{0.9,0.1,0.0}

\renewcommand{\arcsec}{.\hspace{-0.9mm}$^{\prime\prime}$}

\slugcomment{Submitted to ApJS}

\shorttitle{The anisotropic (birefringent) case}
\shortauthors{Bail\'en, Orozco Su\'arez \& Del Toro Iniesta}

\begin{document}

\title{On Fabry-P\'erot etalon-based instruments. III. Instrument applications}

\author{F.J. Bail\'en, D. Orozco Su\'arez, J.C. del Toro Iniesta}
\affil{Instituto de Astrof\'isica de Andaluc\'ia (CSIC), Apdo. de Correos 3004, E-18080 Granada, Spain}
\email{fbailen@iaa.es,orozco@iaa.es,jti@iaa.es}

\begin{abstract}
The spectral, imaging, and polarimetric behavior of Fabry-P\'erot etalons have an influence on imaging vector magnetograph instruments based on these devices. The impact depends, among others, on the optical configuration (collimated or telecentric), on the relative position of the etalon with respect to the polarimeter, on the type of etalon (air-gapped or crystalline), and even on the polarimetric technique to be used (single-beam or dual-beam). In this paper we evaluate the artificial line-of-sight velocities and magnetic field strengths that arise in etalon-based instruments attending to the mentioned factors. We differentiate between signals that are implicit to telecentric mounts due to the wavelength dependence of the point-spread function and those emerging in both collimated and telecentric setups from the polarimetric response of birefringent etalons. For the anisotropic case we consider two possible locations of the etalon, between the modulator and the analyzer or after it, and we include the effect on different channels when dual-beam polarimetry is employed. We also evaluate the impact of the loss of symmetry produced in telecentric mounts due to imperfections in the illumination and/or to a tilt of the etalon relative to the incident beam. 
\end{abstract}


\section{Introduction} \label{sec:intro}
Some solar magnetographs are based on the combination of a polarimeter with a tunable bidimensional filter, typically a Fabry-P\'erot etalon. The final goal of these instruments is to infer with precision the solar magnetic field and plasma velocities from the spectrum and state of polarization of light. Hence, it is mandatory to have control over the polarimetric influence of all optical elements on the polarization measurement process. Usually, the whole system is calibrated in such a way that the Mueller matrix of the instrument contains information on the modulator, the analyzer, and the remaining elements in the optical setup. This way it is not necessary to pay much attention to the polarimetric behavior of the particular optical elements. However, etalons used as monochromators have an impact on the measurement of the Stokes vector even if they are perfectly isotropic. Their influence into real observations is such that it cannot be calibrated using standard techniques (i.e., with flat illumination) and depends on the manner they are illuminated: collimated or telecentric. For a detailed discussion on the imaging performance of etalons in collimated and telecentric configurations, we refer the reader to the following works: \cite{ref:beckers}, \cite{ref:vonderluhe}, \cite{ref:scharmer}, \cite{ref:righini} and \cite{2019ApJS..241....9B}, the first in our series of papers.

In particular, etalons mounted in a telecentric configuration keep (ideally) the same transmission profile across the FoV at the expense of leading to artificial signals in the measured Stokes vector due to asymmetries induced in the point-spread function (PSF) over the spectral profile \citep{ref:beckers}. Moreover, irregularities on the etalon and deviations from perfect telecentric illumination degrade further the measurements.\footnote{We use the term ``perfect telecentrism'' when referring to telecentric illumination in which the chief ray impinges the etalon perpendicularly to its reflecting surfaces. We consider that any deviation from such a situation is an imperfection because it degrades the spectral transmission and the PSF of the instrument. Then, we refer to those cases as ``imperfect telecentrism'', even if the deviation is only caused by a tilt of the etalon while keeping the telecentrism over the FoV.}
 For example, strictly speaking, no PSF can be defined for the system since translational invariance is lost and the spatial response of the etalon is different for each point over the FoV (see Paper I).   Instead, we can only talk of a \emph{local} PSF to stick to simple and known concepts.
 
 Departures of the chief ray from normal incidence produce an asymmetrization of the spectral transmission profile and of the spatial shape of the local PSF of the instrument. It also introduces a widening of the transmission profile and of the local PSF, and a shift of their peaks (Paper I). Defects associated to deviations of the flatness of the reflecting surfaces can also modify the local PSF and the spectral transmission pixel to pixel. 

In collimated setups, the effects associated to fluctuations of the optical path due to roughness errors average over the area of the etalon that is illuminated. In addition, the PSF dependence with wavelength over the passband is nonexistent, but other problems can arise. For instance, we can no longer talk of PSF, much like in the imperfect telecentric case, because of the loss of space invariance associated to a transmission factor which appears in the PSF and that depends on the image plane coordinates (Paper I).  Moreover, the monochromatic transmission can be reduced drastically if defects on the etalon are not kept low enough. 

In both collimated and telecentric configurations the response of the instrument depends on the object itself. Hence, the inferred Stokes vector can be altered simply because of the polychromatic nature of the observations, no matter which configuration is employed. This is of special importance for telecentric mounts, because of the strong spectral dependence their PSFs suffer from. Naturally, changes of the cavity errors during the spectral scan can also have an impact on the measurement of the Stokes vector for both mounts. Such a change on the defects distribution has been confirmed recently in piezo-stabilized etalons \citep{ref:greco}.

Examples of instruments based on etalons illuminated with a telecentric beam are the Italian Panoramic Monochromator at THEMIS \citep[and references therein]{themis}, the TESOS spectrometer at the VTT \citep{kentischer}, the CRisp Imaging SpectroPolarimeter instrument at the Swedish 1 m Solar Telescope \citep{crisp,crisp2},  the PHI instrument on board the Solar Orbiter mission \citep{sophi}, and the Visible Tunable Filter at the upcoming DKIST \citep{ref:schmidt}. Solar instruments equipped with etalons mounted in a collimated setup are the Interferometric BIdimensional Spectrometer (IBIS) at the Dunn Solar Telescope of the Sacramento Peak Observatory \citep{cavallini}, the GFPI at GREGOR \citep{gregor}, and the IMaX instrument aboard {\sc Sunrise} \citep{imax} .

Among the mentioned instruments, IMaX and PHI use Fabry-P\'erots based on lithium niobate crystals to allow for spectral scanning without the need of using piezo-actuators. The birefringent properties of this crystal also contribute to modify the incident Stokes vector. In particular, the polarimetric behavior depends on the etalon geometry, wavelength, angle of the incident wavefront, birefringence of the crystal, and on the orientation of the optical axis angle of the crystal with respect to the wavefront normal in the way described by \cite{2019ApJS..242...21B} (hereafter Paper II). Of course, birefringence can also appear locally within the etalon due to local surface defects created during the polishing and to the polarization-dependent response of the coating of the etalon \citep{ref:doerr}. 

Fortunately, the etalon is never positioned at the beginning of the optical setup when doing full polarimetric measurements. Rather, it is usually illuminated by a polarimetric modulated intensity signal if the etalon is located before the analyzer or just with linearly polarized light when it is at the very end of the optical path, following the analyzer. The influence of the etalon in the polarimetric behavior of polarimeters has been addressed already \citep{ref:jti}. These authors considered the effect of typical optical elements and included a birefringent Fabry-P\'erot, concluding that the optimum polarimetric efficiencies can still be reached no matter the retardance introduced by the etalon. However, they did not take into account either the real Mueller matrix of the etalon nor the influence of the optical configuration, but just represented the etalon as an additional retarder plus a mirror within the optical path.

In a more realistic situation, the birefringent effects brought about by the etalon depend on the optical setup, i.e., on how the etalon is illuminated within the optical path. In collimated setups, the coefficients of the Mueller matrix of the etalon are reduced to four independent terms that vary with the parameters mentioned above (Paper II). The spectral dependence of the coefficients is particularly strong and plays an important role in quasi-monochromatic observations. Moreover, the Mueller matrix shape changes with the principal plane orientation, which is determined by the plane formed by  the wavefront vector and the optical axis of the crystal. This implies that the impact of the birefringence of the etalon is different for each direction of the wavefront and, thus, for each pixel. In perfect telecentric mounts, off-diagonal terms on the Mueller matrix are null and the  effect of the birefringence is translated only into the transmission profile (Paper II). In real instruments in which illumination  differs from perfect telecentrism and/or the local deviations of the optical axis appear during the process of manufacturing (\emph{local domains}),  the Mueller matrix no longer remains diagonal and the effect on the polarimetric measurements is more pronounced, since cross-talks between different Stokes components can appear, just like in the collimated case.

This paper is a continuation of the work presented in \cite{2019ApJS..241....9B,2019ApJS..242...21B} were we reviewed the spectral, imaging, and birefringent properties of Fabry-P\' erot etalons when located in solar magnetographs. Here we evaluate the influence of etalons in the process of measuring physical solar quantities from the observations, that is, we assess their imprints in the inferred line-of-sight (LoS) velocities and the magnetic field strengths from solar vector magnetographs. We begin with an evaluation of artificial signals in isotropic telecentric mounts for both perfect and imperfect illumination of the etalon (Section \ref{sec:artificial_telecentric}). Then, we study the effects of birefringence on the measurements (Section \ref{sec:birefringent}). We consider two possible locations of the etalon: after the polarimeter and between the modulator and the analyzer. We also differentiate between collimated and telecentric setups and we include the effects of imperfect illumination of the etalon. 

\section{Artificial signals in isotropic telecentric mounts}
\label{sec:artificial_telecentric}

\cite{ref:beckers} was the first to predict that the spectral dependence of the PSF implicit to etalons in telecentric configuration gives rise to artificial signals in the LoS velocities. He also warned that these signals are expected to arise in images with velocity structure. The origin of the spurious LoS velocities comes from the fact that observations are not purely monochromatic, but quasi-monochromatic (Eq. [61] of Paper I). The wavelength dependence induces asymmetries in the observed profile even if the original is completely symmetric. Obviously, magnetic field measurements are also influenced by these asymmetries, although not mentioned by \cite{ref:beckers},  and, what is more important, the induced signals cannot be mitigated unless the PSF is completely characterized. The latter is almost impossible in practice.

 A proper evaluation of these false signals requires a careful comparison between a reference case where the spectral
 PSF is assumed to be invariant in wavelength and a real observation in which the PSF varies over the spectral bandwidth. Of course, the modulation scheme of the instrument and the position of the etalon in the optical train must be considered too, specially for etalons that are anisotropic. In this section we will address only the case of etalons that are isotropic (see Sect.~\ref{sec:birefringent} for a discussion on the impact of birefringent etalons). To evaluate the spurious signals, we have compared the expected LoS velocities and magnetic field strength when taking into account the spectral dependence originated in the telecentric case (Paper I) with the results obtained with an \emph{ideal} wavelength independent reference PSF. The ideal PSF we have chosen is simply the monochromatic telecentric PSF at its peak transmission wavelength, modulated spectrally by the  transmission profile that corresponds to the same telecentric configuration. This PSF of reference does not show any spatial variation across the passband of the etalon but only a modulation of intensity, like in a collimated case,  and, hence, it does not introduce spurious signals when measuring the Stokes vector. 

For the test we assume a polarimeter consisting of a pair of liquid crystal variable retarders (LCVRs) as modulator and a linear polarizer as analyzer, similarly to the PHI and IMaX instruments. Application of different voltages to each of the LCVRs translates into different retardances and, consequently, in a modulation of the signal recorded by the instrument cameras. A linear combination of four different modulations is enough to obtain the four Stokes parameters. Furthermore, a proper choice of the retardances optimize the polarimetric efficiencies in the sense of minimizing error propagation in the measurement of the Stokes vector \citep{ref:optimum_modulation}. Table \ref{tab:PM} shows the retardances for both LCVRs, $\delta_1$ and $\delta_2$, and for the four modulations in sequential order (from PM1 to PM4) employed to obtain an optimum modulation scheme in the mentioned instruments. In imaging instruments, this process is done at each of the wavelengths of interest. Once the Stokes parameters have been determined, it is possible, using different diagnostic techniques, to infer the LoS velocity and the vector magnetic field of the plasma.

\begin{table}
	\begin{center}
	\caption{Optimum retardances used for the LCVRs in the IMaX and PHI instruments.}
	\label{tab:PM}
		\begin{tabular}{| l | l | l | l | l |}
			\hline 
			& PM1 & PM2 & PM3 & PM4\\ [0.2cm] \hline
			$\delta_1 [\degree]$ & 225 & 225 & 315 & 315\\ [0.2cm] \hline 
			$\delta_2 [\degree]$ & 234.736 & 125.264 & 54.736 & 305.264 \\[0.2cm]
			\hline
		\end{tabular}
	\end{center}
\end{table}

\begin{figure*}
	\includegraphics[width=\textwidth]{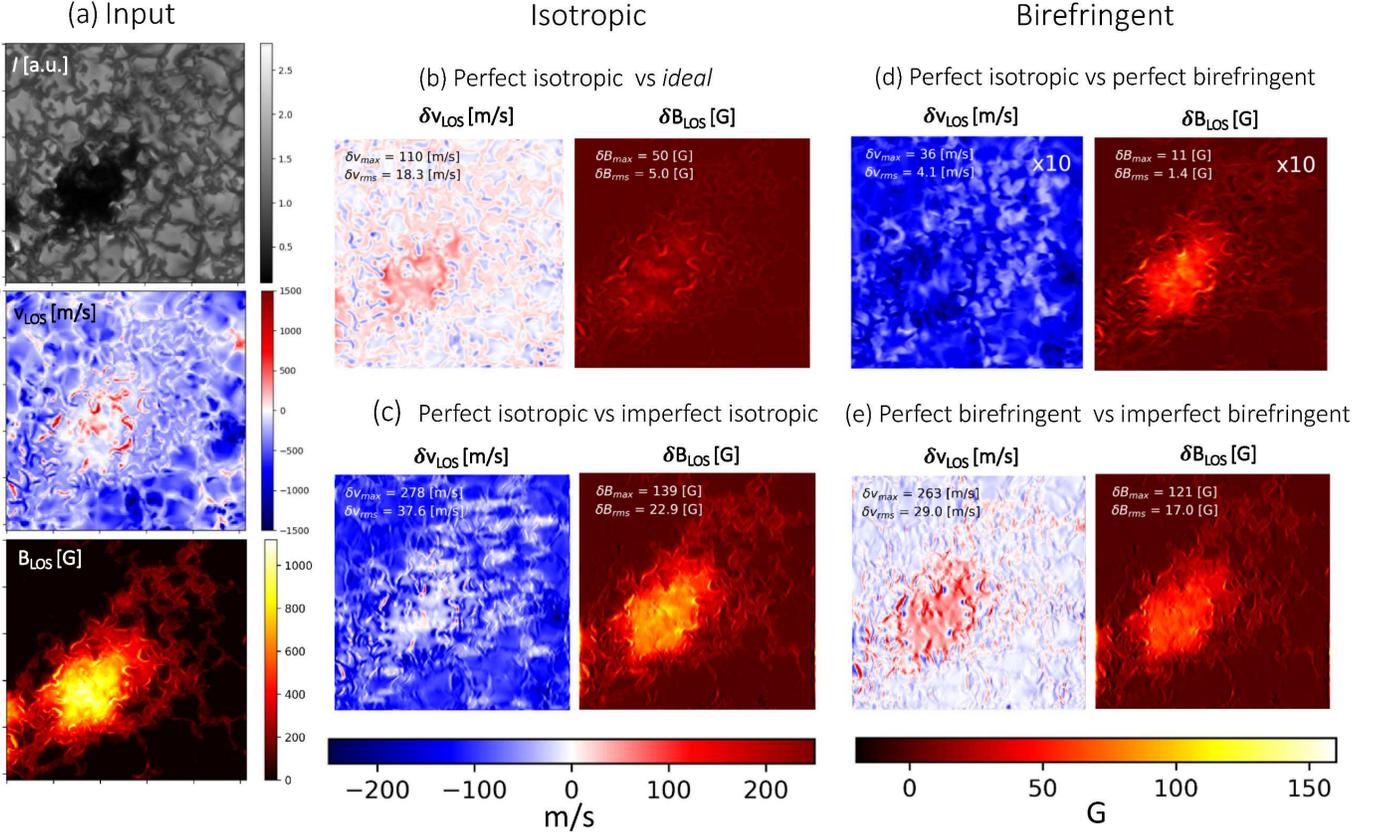}
	\caption{(a) Synthetic input maps from MHD simulations: Stokes $I$  (top); LOS velocities (middle); and LOS magnetic field (bottom). From (b) to (e), comparison of observed LOS velocities (left) and magnetic fields (right) by a telecentric etalon for different situations considering both the isotropic ((b) and (c)) and the birefringent cases ((d) and (e)): (b) residual signal after subtracting the one obtained when employing the reference wavelength independent PSF, labeled as \emph{ideal}, and the signals that appear using the isotropic PSF that considers the wavelength dependence; (c) difference between the ``perfect'' isotropic telecentric configuration where the chief ray is perpendicular to the etalon surfaces with respect to an ``imperfect'' isotropic telecentric configuration in which the chief ray has an incidence angle on the etalon of $0\fdeg 5$; (d) difference between the signals arising for a perfect birefringent and an isotropic mount; (e) same as (c) but considering a birefringent etalon.}
	\label{fig:artificial}
\end{figure*}

We have simulated the effect of a telecentric etalon in the inferred LoS velocities and magnetic field strength on a set of synthetic spectral images of the four Stokes parameters obtained through magnetohydrodynamical (MHD) simulations \citep{2005A&A...429..335V}. The spatial sampling of the synthetic data is 0\arcsec0287 and the size of our image is $256\times 256$ pixels$^2$. The spectral range goes from $-\;40$ pm to $40$ pm in steps of $1$ pm and is centered about the $525.02$ nm Fe~{\small I} line observed by IMaX.\footnote{Although we will concentrate our tests in this spectral line, the results are of the same order for other lines, such as the Fe~{\small I} 617.3 nm observed by PHI.}
 We have modulated the Stokes vector monochromatically with the set of retardances presented in Table~\ref{tab:PM} assuming the etalon is placed after the analyzer. This choice of the etalon position is irrelevant because it is considered to be isotropic, though. Figure \ref{fig:artificial} (a) shows the simulated Stokes $I$ parameter at the continuum (top), as well as the LoS velocity structure (middle) and the magnetic field strength (bottom) corresponding to the input data. A solar pore with an intense magnetic field can be appreciated, covering an area of approximately $100\times 100$ pixels$^2$.  

The different \emph{observed} intensity maps at each wavelength of the spectral range are obtained over the range $\pm \; 20$ pm with respect to the center of the line by tuning the transmission profile of the etalon over the target spectral line and applying Eq. [62] of Paper I. The  PSF considered corresponds to a perfect telecentric $f/40$ isotropic etalon with $n=2.3$, $h=250$ $\mu$m, $R=0.92$, and $A=0$.\footnote{The reader is referred to Paper I for the missing definitions.} Then, we have obtained the Stokes parameters at each wavelength with the proper demodulation matrix \citep{ref:jti}. 
Finally, we have compared the corresponding LoS velocities and magnetic field strengths with the ones obtained with the reference PSF mentioned above.
 
For the sake of simplicity, the LoS velocities and magnetic field strength signals have been calculated using the center of gravity (CoG) method \citep{ref:semel}. Figure \ref{fig:artificial} (b) shows the spurious signals obtained for the LoS velocities (left) and magnetic fields strength (right) in the telecentric case  when compared to the reference case, labeled as \emph{ideal}, defined above. We have restricted also to ``perfect telecentrism'', i.e., to normal incidence of the chief ray on the etalon surfaces for the whole FoV.
  It can be seen that the difference between signals reach values up to $\sim110$ m/s for the LoS velocity,  $\delta v_{\rm LOS}$, and as much as $\sim50$ Gauss for the field strength, $\delta B_{\rm LOS}$, both in absolute value. The artificial LoS velocity map shows considerable small scale fluctuations associated to the presence of granules, intergranular lanes and a pore. Of course, this is because the Stokes parameters have changed after passing through the etalon. Although not shown, it is representative to notice that the artificial signals obtained for Stokes $V$ are always below $\sim5\;\%$ in the wing of the $525.02$ nm Fe~{\small I} line, where the Stokes $V$ reaches a maximum. The rms and maximum values of the spurious signals can be found in Table~\ref{tab:results1}. 

In the case that the cone of rays is inclined a small angle with respect to the normal of the etalon the loss of symmetry with respect to the normal makes both the spectral transmission of the etalon and the spatial PSF to become asymmetric (Paper I). Such an effect happens locally in imperfect telecentric mounts, where the chief ray deviates gradually from the center of the image plane to its borders. It also occurs when the etalon is tilted to suppress  ghost images on the focal plane originated by multiple reflections. Since the effects are equivalent, we will refer hereafter to these two cases indistinctly as ``imperfect telecentrism''.
The induced asymmetries are expected to further introduce false LoS velocities and magnetic field signals.  Naturally, asymmetries in the instrumental profile can arise also from an unsymmetrical spatial distribution of cavity errors \citep[e.g.,][]{ref:reardon}. Figure \ref{fig:artificial} (c)  shows the map of artificial LoS velocities (left) and magnetic signals (right) that appear in an imperfect isotropic telecentric configuration with a chief ray angle of $0\fdeg 5$. As reference, a perfect isotropic telecentric etalon has been considered. Differences between the perfect and imperfect telecentric mounts are as large as $\sim 140$ G in $\delta B_{\rm LOS}$, $\sim280$ m/s in $\delta v_{\rm LOS}$ and $\sim 18\;\%$  in $V$ (Table \ref{tab:results1}). Such high signals are caused by a large shift and a significant asymmetrization of the observed spectral profile. The PSF is shifted and loses its spatial symmetry (see Paper I), thus displacing the profiles and introducing an offset in the velocities ($\sim80$ m/s). The rms value of the artificial velocities is probably better suited for comparison purposes with perfect telecentrism. In this case, the rms velocity is $\sim 37.5$ m/s, whereas for the velocities in Figure ~\ref{fig:artificial} (b) it is approximately half this value, $\sim 18$ m/s. 
Note that typical tolerances in real instruments usually keep deviations below $0\fdeg 5$. Moreover, this value corresponds to a maximum deviation and affects mostly to the borders of the image while here we have assumed that the whole image suffers from such a deviation.

So far, we have restricted to a telecentric configuration with an $f/40$ aperture. Such wide apertures are not usual in solar instruments. Rather, f-numbers are typically larger than $f/100$, specially for ground-based telescopes. Examples are the beams illuminating the etalons of THEMIS ($f/192$), TESOS ($f/125$ and $f/265$) and of the Visible Tunable Filter ($f/200$).  In addition, deviations from perfect telecentrism in these instruments are not as large as the one assumed here. However, tilt of the etalon to suppress ghost images is common and it affects the relative inclination of the cone of rays over the whole FoV. Fortunately, tilts applied are typically far below $0\fdeg5$. PHI is an exception, since its etalon is illuminated by an $f/56$ or by an $f/63$ beam depending on the configuration. In addition,  tolerances in this instrument allow for a maximum deviation of the chief ray over the FoV of $0\fdeg23$. Yet, the impact on the measurements is  more benign than the shown in Sects  \ref{sec:artificial_telecentric}, \ref{sec:after_analyzer_tel} and \ref{sec:before_analyzer_tel} because of the larger aperture and the better telecentrism. 

\begin{figure}
	\begin{center}
		\includegraphics[width=0.48\textwidth]{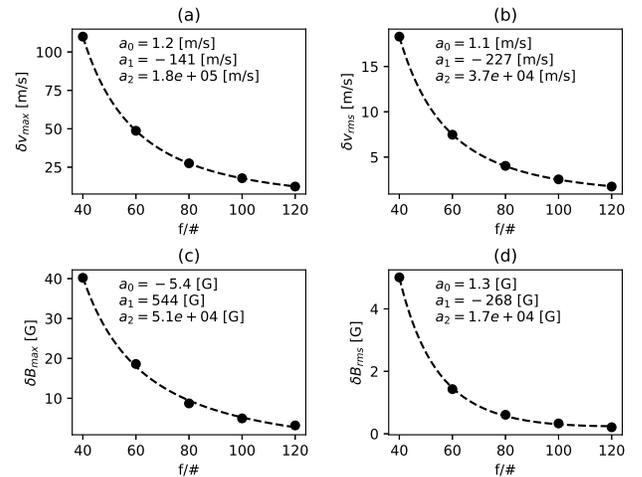}
	\end{center}
	\caption{Maximum and rms value calculated from the maps of artificial LoS velocities ((a) and (b) respectively) and magnetic field ((c) and (d), respectively) arising in a telecentric setup vs the f-number of the beam. Values obtained directly from simulations are displayed as dots, whereas the corresponding fitting is shown as a dashed line.}
	\label{fig:artificial_vs_fnum}
\end{figure}

 Figure \ref{fig:artificial_vs_fnum} shows the maximum and rms values of the spurious signals obtained for a perfect telecentric configuration with f-numbers 40, 60, 80, 100 and 120. Fitting of the data to a curve of the type $a_0 +a_1(f\#)^{-1}+a_2(f\#)^{-2}$ is also displayed for each subfigure, where $a_0$, $a_1$ and $a_2$ are the adjusted coefficients. We observe that artificial signals decay roughly in a quadratic way with the inverse of the $f\#$. Consequently, we can safely disregard the mentioned effects in etalons illuminated by the slow beams associated to ground-based instruments. For PHI, undesired signals are still expected to be seen, although less than half than the ones presented here for an $f/40$ telecentric configuration. The use of such a ``fast'' beam and such a large deviation of the chief ray in our simulations simply serves to illustrate more clearly the possible artificial signals that can appear in telecentric mounts and their appearance. In any case, a careful assessment is required for the future generation of space instruments, which will probably require ``small'' f-numbers (< $f/60$) for compactness purposes.

\section{Effects of etalon birefringence on the polarimetric modulation}
\label{sec:birefringent}

\begin{figure*}[t]
	\centering
	\includegraphics[width=1\textwidth]{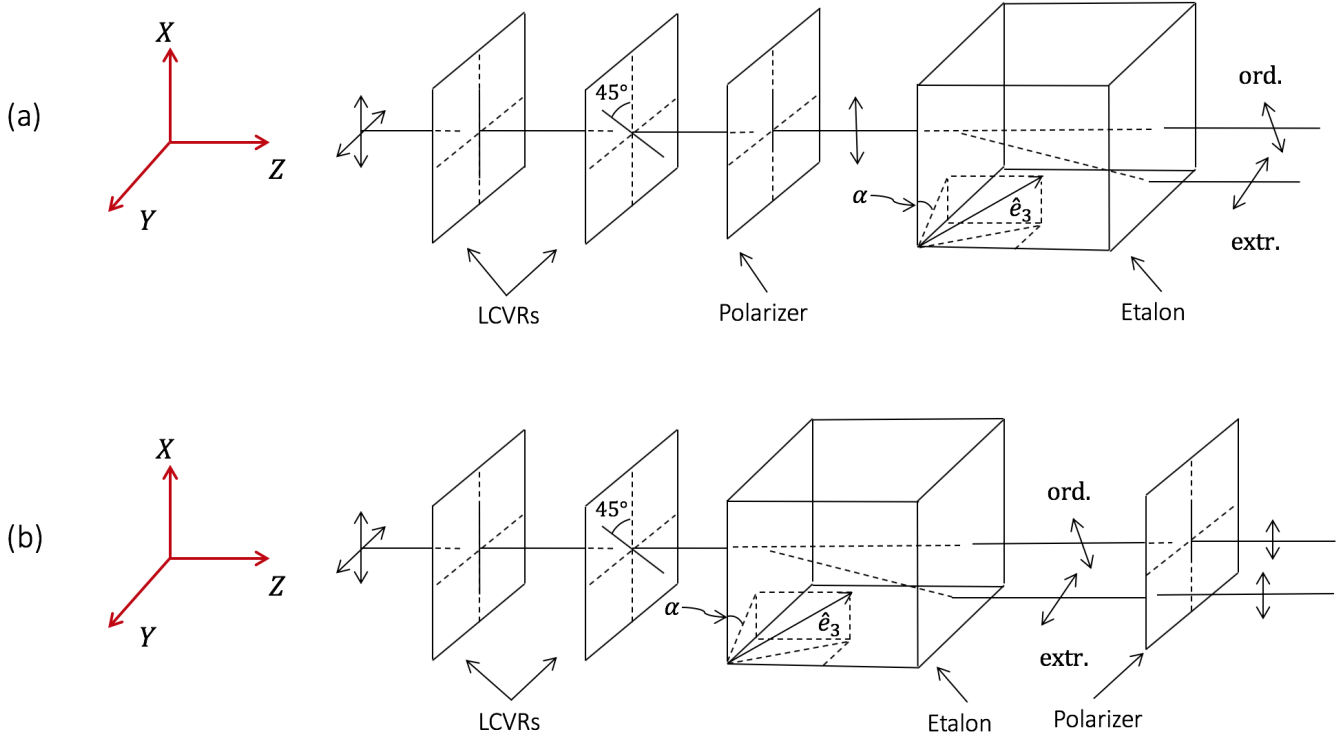}
	\caption{Layout of the transmission of the electric field components of the incident light when: (a) the etalon is located after the analyzer (left); and (b), the etalon is located between the LCVRs and the analyzer (right).}
	\label{fig:scheme}
\end{figure*}

In Paper II we showed that the Mueller matrix of a birefringent etalon is a combination of both a retarder and a mirror modulated by a wavelength-dependent gain factor. Then, any deviation from normal illumination has an impact on the optimum polarimetric efficiencies and on the measured Stokes parameters. The presence of the etalon can be evaluated easily if the polarimetric response of the etalon is included in the Mueller matrix of the polarimeter. A distinction between the next two cases is mandatory: (1)  the etalon is located after the analyzer (Fig.~\ref{fig:scheme} (a)), and (2)  the etalon is placed at an intermediate position between the modulator and the analyzer (Fig.~\ref{fig:scheme} (b)). The second configuration is common in dual-beam polarimeters, such as IMaX, whereas the first is used in single-beam instruments, like PHI. Both use also a birefringent etalon made of lithium niobate.

The illumination of the etalon (collimated or telecentric) is also important in the analysis since it changes the functional shape of the Mueller matrix coefficients. We will consider each case separately in the next sections assuming the same polarimeter as in the previous section.

\begin{table*}
	\begin{center}
		\caption{Summary of results of the artificial signals found in telecentric  ($f/40$) and collimated configuration.}
		\label{tab:results1}
		\vspace{0.2truecm}
		\begin{tabular}{lcccccc}
			\hline 
			{ } \\
			Configuration & $\delta v_{\rm rms}$ [m/s]&$\delta v_{\rm max}$ [m/s] & $\delta B_{\rm rms}$ [G]& $\delta B_{\rm max}$ [G] & $\delta V_{\rm rms}$ [\%]&$\delta V _{\rm max}$ [\%]\\
			{ } \\
			\hline
			\hline
			Isotropic perfect telecentric vs isotropic monochromatic telecentric & 18.3 & 110 & 5.0 & 50 & 0.4 & 4.9\\
			Isotropic imperfect telecentric vs isotropic perfect telecentric & 37.6 & 278 & 22.9 & 139 & 2.2 & 18\\
			Birefringent perfect telecentric vs isotropic perfect telecentric& 4.1 & 36 & 1.4 & 11 & 0.09 & 0.8\\
			Birefringent perfect telecentric (channel 1 vs channel 2)&-&-&-&-&-&0.006\\
			Birefringent imperfect telecentric vs birefringent perfect telecentric& 29.0 & 263 & 17.0 & 121 & 2.3 & 20\\
			Birefringent collimated before analyzer (channel 1 vs channel 2) &-&-&-&-&-&0.3\\
			Birefringent collimated before analyzer vs birefringent collimated after analyzer &3&15&0.07&0.7&0.03&0.45\\	
			\hline
		\end{tabular}
	\end{center}
\end{table*}
\subsection{Etalon located after the analyzer} \label{sec:after_analyzer}

\subsubsection{Collimated configuration}
The Mueller matrix of a polarimeter formed by a pair of LCVRs and an analyzer is given by $\rm {\textbf M}_{\rm pol}=\textbf{LR}_2{\textbf R}_1$, where $\rm \textbf{L}$, $\rm \textbf{R}_2$ and $\rm \textbf{R}_1$ correspond to the Mueller matrices of a linear polarizer with its transmission axis at $0\degree$, a retarder with fast axis at $45\degree$, and a retarder with fast axis at $0\degree$ (all angles measured with respect to the $+\;Q$ direction). The Mueller matrix of the polarimeter can be cast in such a case as

\begin{equation}
{\rm \textbf{M}_{pol}}=\frac{1}{2}\begin{pmatrix}
1 & \cos\delta_2 & \sin\delta_1\sin\delta_2 & -\cos\delta_1\sin\delta_2\\
1 & \cos\delta_2 & \sin\delta_1\sin\delta_2 & -\cos\delta_1\sin\delta_2\\
0 & 0 & 0 & 0\\
0 & 0 & 0 & 0\\
\end{pmatrix},
\label{pmp}
\end{equation}
where $\delta_1$ and $\delta_2$ are the retardances associated to  the LCVRs at $0\degree$ and $45\degree$, respectively.
If we assume that the etalon is in a collimated configuration and that is placed after the analyzer (Fig. \ref{fig:scheme} (a)), then the Mueller matrix of the system is given by $\rm \textbf{M}_{tot}=\textbf{M}_{et}\textbf{M}_{pol}$, where $\rm{\textbf{M}}_{et}$ is the Mueller matrix of the etalon. The Mueller matrix $\rm{\textbf{M}}_{et}$ can be cast as (Eq.~[32] in Paper II)

\begin{equation}
\rm{\textbf{M}}_{et}=\begin{pmatrix}
a & bC_2 & bS_2 & 0\\
bC_2 & aC_2^2+cS_2^2 & (a-c)S_2C_2 & dS_2\\
bS_2 & (a-c)S_2C_2 & aS_2^2+cC_2^2 & -dC_2\\
0 & -dS_2 &dC_2 & c\\
\end{pmatrix},
\label{eq:etalon}
\end{equation}
where $a$, $b$, $c$, and $d$ are defined in Eq. [10] from Paper II and depend on the etalon geometry, wavelength, angle of the
incident wavefront, birefringence of the crystal, and on the
orientation of the optical axis angle of the crystal. Coefficients $C_2\equiv\cos2\alpha$ and $S_2\equiv\sin2\alpha$ arise from a rotation of an angle $\alpha$ about $Z$ that is introduced to take into account the orientation of the etalon principal plane.\footnote{The principal plane is the one formed by the wavefront vector with the optical axis. Its orientation must be taken into account in the analysis because it determines the propagation  properties of orthogonal electric fields.} The multiplication of $\rm{\textbf{M}}_{et}$ by $\rm{\textbf{M}}_{pol}$ yields to

\begin{equation}
{\rm \textbf{M}_{tot}}=\frac{1}{2}\begin{pmatrix}
\Lambda & \Lambda\cos\delta_2 & \Lambda\sin\delta_1\sin\delta_2 & -\Lambda\cos\delta_1\sin\delta_2\\
\Xi & \Xi\cos\delta_2 & \Xi\sin\delta_1\sin\delta_2 & -\Xi\cos\delta_1\sin\delta_2\\
\Pi & \Pi\cos\delta_2 & \Pi\sin\delta_1\sin\delta_2 & -\Pi\cos\delta_1\sin\delta_2\\
\Sigma & \Sigma\cos\delta_2 & \Sigma\sin\delta_1\sin\delta_2 & -\Sigma\cos\delta_1\sin\delta_2\\
\end{pmatrix},
\label{eq:Mtot}
\end{equation}
where
\begin{equation}
\begin{gathered}
\Lambda=a+bC_2,\\
\Xi=bC_2+aC_2^2+cS_2^2,\\
\Pi=bS_2+(a-c)C_2S_2,\\
\Sigma=-dS_2.\\
\end{gathered}
\end{equation}
Then, the instrument modulation matrix is given by
\begin{equation}
\begin{gathered}
{\rm O}=g^{(+)}(\lambda)\begin{pmatrix}
1 & \cos\delta_2^{(1)} & \sin\delta_1^{(1)}\sin\delta_2^{(1)} & -\cos\delta_1^{(1)}\sin\delta_2^{(1)}\\
1 & \cos\delta_2^{(2)} & \sin\delta_1^{(2)}\sin\delta_2^{(2)} & -\cos\delta_1^{(2)}\sin\delta_2^{(2)}\\
1 & \cos\delta_2^{(3)} & \sin\delta_1^{(3)}\sin\delta_2^{(3)} & -\cos\delta_1^{(3)}\sin\delta_2^{(3)}\\
1 & \cos\delta_2^{(4)} & \sin\delta_1^{(4)}\sin\delta_2^{(4)} & -\cos\delta_1^{(4)}\sin\delta_2^{(4)}\\
\end{pmatrix},
\label{eq:Otot}
\end{gathered}
\end{equation} 
where 
\begin{equation}
g^{(+)}(\lambda)=\frac{a(\lambda)+b(\lambda)\cos2\alpha}{2}.
\end{equation}
The superscript in $\delta_1$ and $\delta_2$ enumerates the sequential order of the modulation, i.e, $\delta^{(1)}$ corresponds to the retardance for modulation PM1 in Table \ref{tab:PM}, $\delta_1^{(2)}$ refers to modulation PM2, etc. The superscript $(+)$ has been introduced to emphasize that the etalon is illuminated with linear polarization along the $+\;Q$ direction.

The modulation scheme is the same as that of a polarimeter in which the presence of the etalon is neglected, except for a gain factor that depends on both the wavelength and on the direction of the wavefront normal. This factor also varies across the etalon whether the illumination is not homogeneous or the optical axis is deviated from the $Z$ direction, which occurs in local domains, i.e., in regions that suffer from  local imperfections that change the crystal optical axis. In any case, the gain factor is absorbed in what is known as ``flat field" of the instrument, a correction factor that takes into account inhomogeneities in the distribution of intensity on the detector because of local changes in the transmission.
 Therefore, the modulation scheme of Table~\ref{tab:PM} remains optimum at each particular monochromatic wavelength even when considering the birefringence of the etalon. Also note that the PSF in this configuration is the same to that of an ideal circular aperture modulated by the transmission factor $g^{(+)}(\lambda)$. Hence, \emph{the measured Stokes parameters are expected to be insensitive to birefringence whenever the etalon is positioned after the polarimeter and illuminated by a collimated beam.}

\subsubsection{Telecentric configuration}
\label{sec:after_analyzer_tel}
Let us assume now that the etalon is not illuminated with collimated light but, instead, with a telecentric beam. In this configuration, Eq.~(\ref{eq:etalon}) does not hold and we need to use Eq.~(48) from Paper II. Therefore, the Mueller matrix of the etalon, ${\rm \tilde{\textbf{M}}_{et}}$, is now given by
\begin{equation}
{\rm \tilde{\textbf{M}}_{et}}=
\begin{pmatrix}
\tilde{a}\prima & \tilde{b}\prima & 0 & 0\\
\tilde{b}\prima & \tilde{a}\prima & 0 & 0\\
0 & 0 & \tilde{c}\prima & -\tilde{d}\prima\\
0 & 0 & \tilde{d}\prima& \tilde{c}\prima
\end{pmatrix},
\label{Mtele}
\end{equation}
where coefficients $\tilde{a}\prima$, $\tilde{b}\prima$,  $\tilde{c}\prima$, and  $\tilde{d}\prima$ are defined in Eq.~[49] of Paper II.\footnote{Note that tildes are employed to allude to the telecentric configuration. This notation is consistent with that of Paper II.} These coefficients vary in a different manner when compared to the collimated case with the wavelength, etalon geometry, birefringence, etc. 
 The modulation matrix remains the same to Eq.~(\ref{eq:Otot}), except for the gain factor, which is given in this case by

\begin{equation}
\tilde{g}^{(+)}(\lambda)=\frac{\tilde{a}\prima(\lambda)+\tilde{b}\prima(\lambda)}{2}.
\label{eq:gplus}
\end{equation}
Hence, the modulation scheme of Table \ref{tab:PM} remains optimum monochromatically in a telecentric birefringent configuration, as for collimated setups. However, the PSF is different compared to that of the isotropic telecentric configuration. In particular, an asymmetry on the spatial shape of the ``birefringent'' PSF is induced along two perpendicular directions  even for perfect telecentrism, i.e., it becomes elliptic. This is shown in Appendix \ref{app:orthogonal_psfs}. Differences between the isotropic and the birefringent PSFs can be interpreted as spurious signals in the measured Stokes parameters that must be added to those presented in previous sections.  Figure~\ref{fig:artificial} (d) shows the artificial LoS velocities (left) and magnetic field strength (right) when comparing the telecentric isotropic case against the telecentric birefringent case. Note that the maps have been multiplied by a factor $\times 10$ to maintain the same color scale in all subfigures from (b) to (e). This means that signals are about an order the magnitude lower than the ones obtained in the other cases. In particular, the maximum difference at the wing of the line in $V$ is $\sim 0.8\;\%$ and about $10$ G and $35$ m/s in the LoS magnetic field and velocities (Table~\ref{tab:results1}).\footnote{We have employed ordinary and extraordinary refraction indices $n_o=2.3$ and $n_e=2.2$, which corresponds to lithium niobate, for the simulations of the birefringent Fabry-P\'erot. The remaining parameters of the etalon are the same as in previous sections and are consistent with the simulations presented in Paper II.} 

Obviously, deviations of the chief ray angle from normal illumination can also contribute to the emergence of artificial signals, as in the isotropic case. Figure \ref{fig:artificial} (e) shows the difference between the observed LoS velocities and magnetic field strength compared to the perfect birefringent telecentric case. Differences in the magnetic field are as much as $120$ G and $260$ m/s for the LOS velocities (Table~\ref{tab:results1}). The maximum value of the artificial $V$ at the wing of the line is about $20\;\%$. Note that the results are comparable to those obtained for the isotropic case in Figure \ref{fig:artificial} (c), which indicates that the impact on the measurements due to the anisotropy of the etalon is small compared to the effect of the wavelength dependence of the PSF intrinsic to these mounts whether the Fabry-P\'erot is birefringent or not.

\subsection{Etalon located between the modulator and the analyzer}
\label{sec:before_analyzer}
\subsubsection{Collimated configuration}
\label{sec:befor_analyzer_collimated}
In dual-beam instruments the etalon is never placed after the polarimeter. Instead,  it is located between the modulator and the analyzer (Fig.~\ref{fig:scheme} (b)) to avoid the use of two etalons, one for each orthogonal beam in which light is split. A good example of a dual-beam instrument is IMaX, which employs a beam splitter as analyzer to record orthogonal polarizations, $\pm\;Q$, in two different cameras. In this type of configuration etalon anisotropies are expected to have a stronger impact on the measured Stokes vector than when located after the polarizer and, thus,  illuminated with linear polarization. In particular, the optimum modulation scheme presented in Table \ref{tab:PM} can no longer be optimum \citep{ref:jti} and the measured Stokes vector can differ for orthogonal channels only because of the birefringence of the etalon. 

 Assuming that the etalon is in a collimated configuration and following the notation of Section \ref{sec:after_analyzer}, the Mueller matrix is given by $\rm \textbf{M}_{\rm pol}= \textbf{LF}$, where we have defined $\rm \textbf{F}$ as ${\rm\textbf{M}_{et}}\textbf{R}_2\textbf{R}_1$. Given that only the coefficients of the first two rows and columns of $\rm \textbf{L}$ are different from zero, we just need to calculate the coefficients of the first two rows of $\rm \textbf{F}$ to derive the modulation matrix of the instrument: 
\begin{equation}
\begin{gathered}
{\rm F}_{11}=a, \\[3pt]
{\rm F}_{12}=bC_2\cos\delta_2,\\[3pt]
{\rm F}_{13}=b\sin\delta_2(C_2\sin\delta_1-S_2\cos\delta_1),\\[3pt]
{\rm F}_{14}=b(S_2\sin\delta_1-C_2\sin\delta_2\cos\delta_1),\\[3pt]
{\rm F}_{21}=bC_2,\\[3pt]
{\rm F}_{22}=(aC_2^2+cS_2^2)\cos\delta_2+dS_2\sin\delta_2,\\[3pt]
{\rm F}_{23}=(aC_2^2+cS_2^2)\sin\delta_2\sin\delta_1+(a-c)S_2C_2\cos\delta_1\\
-dS_2\sin\delta_1\cos\delta_2,\\[3pt]
{\rm F}_{24}=-(aC_2^2+cS_2^2)\sin\delta_2\cos\delta_1+(a-c)S_2C_2\sin\delta_1\\
+dS_2\cos\delta_1\cos\delta_2.
\end{gathered}
\end{equation} 
When dual-beam techniques are employed, we must differentiate between the Mueller matrices corresponding to the $\pm\;Q$ channels. For the $+\;Q$ channel:
\begin{equation}
{\rm \textbf{M}_{pol}^{(+)}}=\frac{1}{2}\begin{pmatrix}
F_{11}+F_{21} & F_{12}+F_{22} & F_{13}+F_{23} &F_{14}+F_{24}\\
F_{11}+F_{21} & F_{12}+F_{22} & F_{13}+F_{23} &F_{14}+F_{24}\\
0 & 0 & 0 & 0\\
0 & 0 & 0 & 0\\
\end{pmatrix}.
\label{Mtot2}
\end{equation}
For the $-\;Q$ channel, the Mueller matrix is just
\begin{equation}
{\rm \textbf{M}_{pol}^{(-)}}=\frac{1}{2}\begin{pmatrix}
F_{11}-F_{21} & F_{12}-F_{22} & F_{13}-F_{23} & F_{14}-F_{24}\\
F_{21}-F_{11} & F_{22}-F_{12} & F_{23}-F_{13} & F_{24}-F_{14}\\
0 & 0 & 0 & 0\\
0 & 0 & 0 & 0\\
\end{pmatrix}.
\label{Mtot2minus}
\end{equation}

Each row of modulation matrix, $\rm \textbf{O}$ corresponds to the first row of Eqs.~(\ref{Mtot2}) or (\ref{Mtot2minus}) evaluated for the particular retardances of the LCVRs of the modulation scheme. Note that in this case the optimum modulation scheme depends on $a$, $b$, $c$, $d$, on the channel ($\pm\;Q$), and on the orientation of the principal plane of light. Hence, it differs in general from the one showed in Table \ref{tab:PM} and varies over the FoV for each monochromatic wavelength. 

  \begin{figure}[t]
	\begin{center}
		\includegraphics[width=0.48\textwidth]{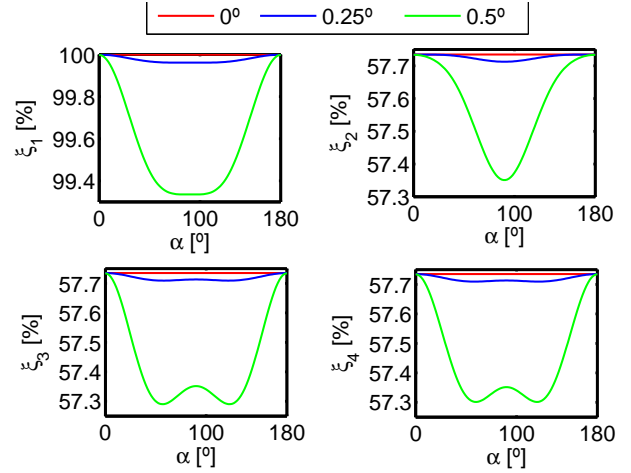}
	\end{center}
	\caption{Components of the efficiency vector as a function of the orientation of principal plane of the etalon using the modulation scheme of Table \ref{pmp} for illumination of the etalon with incident angles $\theta_i=0\degree$ (red solid line), $\theta_i=0.25\degree$ (blue solid line), and $\theta_i=0.5\degree$ (green solid line). The wavelengths at which the transmission profile peaks, $\lambda_{\rm p}$, have been employed at each incident angle. The etalon is located between the modulators and the analyzer in this configuration.}
	\label{eff}
\end{figure}

 Figure \ref{eff} shows the dependence of the efficiency vector \citep{ref:collados} for the $+\;Q$ channel as a function of the orientation of the principal plane, $\alpha$, when using the modulation scheme of Table \ref{tab:PM}. Results are shown for incident angles $\theta=0\degree$, $0\fdeg25$, and $0\fdeg5$ at the corresponding peak wavelengths of the transmission profile $\lambda_{\rm p}=\lambda_0+\Delta\lambda_1$, $\lambda_0+\Delta\lambda_2$, and $\lambda_0+\Delta\lambda_3$, where $\Delta\lambda_1=0$ pm, $\Delta\lambda_2=-\;1.18$ pm, and $\Delta\lambda_3=-\;4.54$ pm. It can be observed that the efficiency decreases from the optimum value whenever $\alpha\neq 0\degree, 180\degree$. The maximum variation is $\sim0.6\;\%$ for the first component of the efficiency vector  and $\sim0.4\;\%$ for the other components.

\begin{figure}[t]
	\begin{center}
		\includegraphics[width=0.52\textwidth]{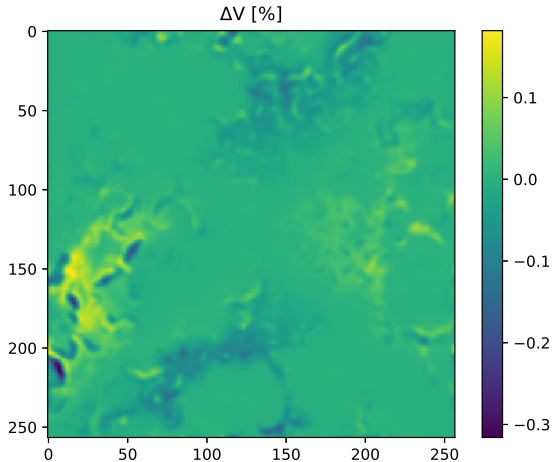}
	\end{center}
	\caption{Difference between the measured circular polarization in each channel at the wing of the Fe I 525 line for an anisotropic (uniaxial) collimated etalon placed between the LCVRs and the analyzer.}
	\label{fig:artificial_V_coll}
\end{figure}

Demodulation with such a non-optimum scheme in a collimated etalon can introduce further artificial signals in the measured Stokes parameters than the presented above. Moreover, the spurious signals are different for the two orthogonal beams. Figure \ref{fig:artificial_V_coll} shows the map with the difference between the measured Stokes $V$ at its wing on the $\pm,\;Q$ channels for a collimated etalon with maximum incidence angle $0\fdeg 5$ (which corresponds to the outermost parts of the FoV). In IMaX, the maximum incidence angle is $0\fdeg 44$. Differences are below $\sim0.3\;\%$ in our simulations, so we can safely disregard this effect in that instrument. 

We have also compared the measured LoS velocities and magnetic field strengths for a collimated configuration that uses dual beam with respect to another where the etalon is placed after the analyzer. For the dual beam configuration, the Stokes parameters have been obtained by averaging the signals recorded at each channel. The rms difference in magnetic field strength is below $0.7$ Gauss. Velocities differ less than $15 $ m/s and the maximum artificial signal in $V$ is $\sim 0.45\;\%$.  

\subsubsection{Telecentric configuration}
\label{sec:before_analyzer_tel}
When the etalon is mounted on a telecentric configuration, its Mueller matrix becomes particularly symmetric, as shown in Equation~(\ref{Mtele}). In fact, the Mueller matrix of the etalon commutes  with that of the analyzer due to their symmetry. It is equivalent, then, to place the Fabry-P\'erot either before or after the analyzer. The only difference being that dual-beam techniques can be used only if placed before the analyzer. In that case, the modulation matrix for the $+Q$ channel is given by Eq.~(\ref{eq:Otot}) except for the gain factor,  determined by Eq.~(\ref{eq:gplus}), as explained in Section \ref{sec:after_analyzer_tel}. The gain factor corresponding to modulation matrix of the $-\;Q$ channel is the same except for a minus sign that changes $\tilde{a}\prima+\tilde{b}\prima$ by $\tilde{a}\prima-\tilde{b}\prima$. 

Although monochromatic polarimetric efficiencies remain optimum, the measured Stokes parameters in each channel are expected to be somewhat different because the PSFs change slightly for orthogonal polarizations (Paper II). This effect induces cross-talk signals in the measured Stokes parameters.  We have obtained that the maximum difference between both channels is $\sim 0.006\%$ in $V$ at the wing of the line. This additional contribution to the spurious signals is insignificant compared to the previous ones and it naturally disappears in isotropic etalons.

\section{Summary and conclusions}
An evaluation of the artificial LoS velocities and magnetic field strength signals that arise in magnetographs based on Fabry-P\'erot etalons has been performed. We have distinguished between telecentric and collimated illumination of both crystalline an isotropic etalons. We have also considered different locations of the etalon within the optical path, in particular instruments where the etalon is placed after the polarimeter and those in which it is positioned in an intermediate location between the modulator and the analyzer to allow for dual-beam polarimetry. Our analysis has consisted in simulating the impact of an etalon-based instrument similar to IMaX and PHI on the maps of the Stokes components along the 525.02 nm Fe~{\small I} Zeeman sensitive line. 

For a telecentric $f/40$ isotropic case, spurious velocities obtained through the CoG method are as large as 110 m/s, whereas the magnetic field and Stokes $V$ reach values up to 50 G and $5\;\%$, respectively. These signals are originated by a severe dependence of the PSF shape with the wavelength across the transmission profile. In telecentric mounts affected by a departure of the chief ray of $0\fdeg5$, signals can be as high as 280 m/s and 140 G for the LoS velocities and magnetic field strength when compared to the telecentric isotropic configuration. A shift in the map of velocities arises also because of an asymmetrization of the  transmission profile and of the PSF. Apart from the shift, the map of artificial velocities shows structures with a corresponding rms value twice as large as for the perfect configuration ($\sim 37.5$ m/s).  We have also shown that the expected spurious signals in telecentric configurations mounted in ground-based instruments are almost insignificant because of the very slow apertures employed in such telescopes. In particular, we expect a decrease of the spurious signals with $\sim (f\#)^{-2}$. Attention must be paid in etalons aboard space instruments, though, because size constraints usually lead to much faster apertures than the ones typical to ground-based instruments. Collimated isotropic setups are exempt from the emergence of the artificial signals that occur in telecentric mounts, since no spectral variation of the PSF appears in this configuration.

In relation to birefringent etalons, we have showed that the ideal modulation scheme derived by \cite{ref:jti} still remains optimum for both telecentric and collimated setups regardless of the birefringence exhibited by the Fabry-P\'erot as long as the etalon is placed after the polarimeter. We have proved that the measurement of the Stokes parameters is also insensitive to birefringence  in the case of collimated etalons positioned after the analyzer. The same does not apply in telecentric setups, where the PSF differs from the isotropic case and becomes elliptic. Compared to the isotropic telecentric case, artificial signals in the velocities and magnetic field are below 40 m/s and 9.5 G, respectively. The Stokes $V$ is 0.8 \% at most in the wings of the profile .These artificial signals are an order of magnitude smaller than the ones produced simply because of the wavelength dependence of the PSF. Deviations from perfect telecentrism have a similar impact to the one discussed for the isotropic case. 
 
 In a telecentric configuration, placing the etalon between the modulator and the analyzer has the same impact as locating it after the polarimeter since its Mueller matrix commutes with that of the analyzer. Moreover, cross-talks between orthogonal channels are negligible when using dual-beam techniques.  For the collimated configuration, the Mueller matrix of the polarimeter is modified when the etalon is situated before the analyzer and the optimum efficiencies are affected. We have shown that the efficiencies depend on the incident wavefront direction and on the wavelength. However, when the optimum modulation scheme is employed, monochromatic efficiencies decrease only 0.6 \% at most for a $0\fdeg5$ incidence. The measured Stokes vector is also different from the one corresponding to an etalon located after the polarimeter, although differences are below 15 m/s for the velocity, 0.7 G for the magnetic field and 0.45 \% for Stokes $V$ compared to a collimated setup in which the etalon is placed after the analyzer. For this configuration, signals recorded by orthogonal channels in dual-beam instruments are also different due to the presence of the etalon, but these are kept below 0.3 \% at $V$ Stokes wing.

\begin{acknowledgements}
This work has been supported by the Spanish Ministry of Economy and Competitiveness through projects ESP2014-56169-C6-1-R  and ESP-2016-77548-C5-1-R and by Spanish Science Ministry ``Centro de Excelencia Severo Ochoa'' Program under grant SEV-2017-0709 and project RTI2018-096886-B-C51. D.O.S. also acknowledges financial support through the Ram\'on y Cajal fellowship.
\end{acknowledgements}

\appendix
\section{PSF in orthogonal directions: birefringent case}
\label{app:orthogonal_psfs}
Anisotropies in the etalon cause an asymmetry of the PSF on orthogonal directions even if telecentrism is perfect (and, hence, the Jones matrix terms only depend on the radial coordinates of the pupil). Let us consider that the etalon is illuminated with Stokes components $I=Q$ and $U=V=0$ (as for channel $+\;Q$). According to Paper II, the PSF is then given by $\PSF=\atilde\prima+\btilde\prima=\Htildeel\prima_{11}\Htildeel^{\prime\ast}_{11}$. For a perfect telecentric configuration, it was seen in Paper II that the first Jones coefficient are given by
\begin{equation}
\Htildeel\prima_{11}=\int_{0}^{R_{\rm p}}\!\!\!\!\!\int_{0}^{2\pi}\!\!\!\!\!\!r\left[\H_{11}(r,\lambda)\cos^2\phi+ \H_{22}(r,\lambda)\sin^2\phi\right]{\rm e}^{-{\rm i} kr(\alpha \cos\phi + \beta \sin\phi)} \, {\rm d}r \, {\rm d}\phi.
\label{exactH}
\end{equation}

Since $\H_{11}$ and $\H_{22}$ only depend on the radial coordinate of the pupil, we can cast this integral as
\begin{equation}
\Htildeel\prima_{11}=\int_{0}^{R_{\rm p}}\!\!\!\!\!r\H_{11}(r,\lambda)\int_{0}^{2\pi}\!\!\!\!\!\!\cos^2\!\phi \,\,{\rm e}^{-{\rm i} kr(\alpha \cos\phi + \beta \sin\phi)} \, {\rm d}r \, {\rm d}\phi+\int_{0}^{R_{\rm p}}\!\!\!\!\!r\H_{22}(r,\lambda)\int_{0}^{2\pi}\!\!\!\!\!\!\sin^2\!\phi \,\,{\rm e}^{-{\rm i} kr(\alpha \cos\phi + \beta \sin\phi)} \, {\rm d}r \, {\rm d}\phi.
\label{exactH_2}
\end{equation}

Let us take into account two orthogonal directions in the image plane. For example, the direction along $\xi$ and the direction along $\eta$. The Jones term for each case is just

\begin{equation}
\Htildeel\prima_{11}(\xi,\eta=0)=\int_{0}^{R_{\rm p}}\!\!\!\!\!r\H_{11}(r,\lambda)\int_{0}^{2\pi}\!\!\!\!\!\!\cos^2\!\phi \,\,{\rm e}^{-{\rm i} kr\alpha \cos\phi} \, {\rm d}r \, {\rm d}\phi+\int_{0}^{R_{\rm p}}\!\!\!\!\!r\H_{22}(r,\lambda)\int_{0}^{2\pi}\!\!\!\!\!\!\sin^2\!\phi \,\,{\rm e}^{-{\rm i} kr\alpha \cos\phi} \, {\rm d}r \, {\rm d}\phi.
\label{eq:H11_xi}
\end{equation}

\begin{equation}
\Htildeel\prima_{11}(\xi=0,\eta)=\int_{0}^{R_{\rm p}}\!\!\!\!\!r\H_{11}(r,\lambda)\int_{0}^{2\pi}\!\!\!\!\!\!\cos^2\!\phi \,\,{\rm e}^{-{\rm i} kr\beta \sin\phi} \, {\rm d}r \, {\rm d}\phi+\int_{0}^{R_{\rm p}}\!\!\!\!\!r\H_{22}(r,\lambda)\int_{0}^{2\pi}\!\!\!\!\!\!\sin^2\!\phi \,\,{\rm e}^{-{\rm i} kr\beta \sin\phi} \, {\rm d}r \, {\rm d}\phi.
\label{eq:H11_eta}
\end{equation}

The two integrals differ from the exponent of the complex exponential. It turns out that

\begin{equation}
\int_{0}^{2\pi}\!\!\!\!\!\!\cos^2\!\phi \,\,{\rm e}^{-{\rm i} kr\alpha \cos\phi} \, {\rm d}r \, {\rm d}\phi=\int_{0}^{2\pi}\!\!\!\!\!\!\sin^2\!\phi \,\,{\rm e}^{-{\rm i} kr\beta \sin\phi} \, {\rm d}r \, {\rm d}\phi,
\label{eq:equality1}
\end{equation}
and
\begin{equation}
\int_{0}^{2\pi}\!\!\!\!\!\!\sin^2\!\phi \,\,{\rm e}^{-{\rm i} kr\alpha \cos\phi} \, {\rm d}r \, {\rm d}\phi=\int_{0}^{2\pi}\!\!\!\!\!\!\cos^2\!\phi \,\,{\rm e}^{-{\rm i} kr\beta \sin\phi} \, {\rm d}r \, {\rm d}\phi.
\label{eq:equality2}
\end{equation}
Similarly, for the second diagonal element of the Jones matrix
\begin{equation}
\Htildeel\prima_{22}=\int_{0}^{R_{\rm p}}\!\!\!\!\!r\H_{22}(r,\lambda)\int_{0}^{2\pi}\!\!\!\!\!\!\cos^2\!\phi \,\,{\rm e}^{-{\rm i} kr(\alpha \cos\phi + \beta \sin\phi)} \, {\rm d}r \, {\rm d}\phi+\int_{0}^{R_{\rm p}}\!\!\!\!\!r\H_{11}(r,\lambda)\int_{0}^{2\pi}\!\!\!\!\!\!\sin^2\!\phi \,\,{\rm e}^{-{\rm i} kr(\alpha \cos\phi + \beta \sin\phi)} \, {\rm d}r \, {\rm d}\phi.
\end{equation}

Hence,
\begin{equation}
\Htildeel\prima_{22}(\xi,\eta=0)=\int_{0}^{R_{\rm p}}\!\!\!\!\!r\H_{22}(r,\lambda)\int_{0}^{2\pi}\!\!\!\!\!\!\cos^2\!\phi \,\,{\rm e}^{-{\rm i} kr\alpha \cos\phi} \, {\rm d}r \, {\rm d}\phi+\int_{0}^{R_{\rm p}}\!\!\!\!\!r\H_{11}(r,\lambda)\int_{0}^{2\pi}\!\!\!\!\!\!\sin^2\!\phi \,\,{\rm e}^{-{\rm i} kr\alpha \cos\phi} \, {\rm d}r \, {\rm d}\phi.
\end{equation}

\begin{equation}
\Htildeel\prima_{22}(\xi=0,\eta)=\int_{0}^{R_{\rm p}}\!\!\!\!\!r\H_{22}(r,\lambda)\int_{0}^{2\pi}\!\!\!\!\!\!\cos^2\!\phi \,\,{\rm e}^{-{\rm i} kr\beta \sin\phi} \, {\rm d}r \, {\rm d}\phi+\int_{0}^{R_{\rm p}}\!\!\!\!\!r\H_{11}(r,\lambda)\int_{0}^{2\pi}\!\!\!\!\!\!\sin^2\!\phi 
\end{equation}

Using Eqs.~(\ref{eq:equality1}) and (\ref{eq:equality2}), we can see that

\begin{equation}
\begin {gathered}
\Htildeel\prima_{11}(\xi,\eta=0)=\Htildeel\prima_{22}(\xi=0,\eta),\\
\Htildeel\prima_{11}(\xi=0,\eta)=\Htildeel\prima_{22}(\xi,\eta=0).
\end{gathered}
\end{equation}

That is, if $\PSF=\atilde\prima+\btilde\prima$, then $\atilde\prima(\xi,\eta=0)=\atilde\prima(\xi=0,\eta)$, but $\btilde\prima(\xi,\eta=0)=-\btilde\prima(\xi=0,\eta)$ and, therefore, $\PSF(\xi,\eta=0)\neq\PSF(\xi=0,\eta)$. In consequence, the PSF varies for orthogonal directions in birefringent etalons and the symmetry of the PSF is no longer preserved. In practice, the loss of spatial symmetry has a low impact on the measurements since $\btilde$ is much smaller than $\atilde$ (Paper II), as shown in Section \ref{sec:after_analyzer_tel}.


\end{document}